\documentclass[aps,prd,onecolumn,superscriptaddress,nofootinbib,amsmath,amssymb]{revtex4}

\usepackage{graphicx}
\usepackage{wrapfig}

 \tighten  \usepackage[hypertexnames=false]{hyperref}
\begin{document}

 \title{The bigravity black hole and its thermodynamics}

\author{M\'{a}ximo Ba\~{n}ados}
\email{maxbanados@fis.puc.cl}
\affiliation{Departamento de F\'{\i}sica,\\
P. Universidad Cat\'{o}lica de Chile, Casilla 306, Santiago 22,Chile. }

\author{Andr\'es Gomberoff}
\email{agomberoff@unab.cl}
\affiliation{Departamento de Ciencias F\'{\i}sicas,\\
 Universidad Andres Bello, Av. Rep\'{u}blica 252, Santiago,Chile. }

\author{Miguel Pino}
\email{mnpino@uc.cl}
\affiliation{Departamento de F\'{\i}sica,\\
P. Universidad Cat\'{o}lica de Chile, Casilla 306, Santiago 22,Chile. }

\begin{abstract}
We argue that the Isham-Storey exact solution to bigravity does not describe black holes because the horizon is a singular surface. However, this is not a generic property of bigravity, but a property of a particular potential.  More general potentials do accept regular black holes. For regular black holes, we compute the total energy and thermodynamical parameters. Phase transitions occur for certain critical temperatures. We also find a novel region on phase space describing up to 4 allowed states for a given temperature.
\end{abstract}

\maketitle

\section{The bigravity action}

Bigravity is a theory of gravity with two independent {\it dynamical} metrics (denoted  by $g_{\mu\nu}$ and $f_{\mu\nu}$). The action, first considered by Isham-Salam-Strathdee \cite{ISS}, is
\begin{eqnarray}\label{I}
I[g_{\mu\nu},f_{\mu\nu}] &=& {1 \over 16\pi G} \int \Big[\ \sqrt{g} R(g) + \sigma\, \sqrt{f} R(f)- U(g,f) \Big]
\end{eqnarray}
where $U(g,f)$ is an interaction potential depending both on $g$ and $f$. The dimensionless parameter $\sigma$ measures the relative strengths of both Newton's constants.

Bigravity has received intermittent but consistent attention since it was first presented. For some recent work see \cite{DamourKogan,Arkani-Hamed2003,BlasDeffayetGarriga,Berezhiani:2007zf,Blas:2008uz}. The first question one may ask is what is the physical metric determining the geometry of spacetime. This is the same as asking to what metric do particles couple to. We refer the reader to the literature for discussions on many different interpretations. In this work, we shall work entirely on vacuum space filled only by the two metrics. The problem we shall be concerned --regularity of the black hole horizon-- is independent on the interpretation for the bigravity theory.

The potential considered in \cite{ISS} was \footnote{In \cite{ISS} $\sqrt{f}$ is replaced by $|g|^{v} |f|^{1/2-v}$ which is consistent with diff invariance. We have set $v=0$ for simplicity, and to make contact with \cite{BGRS,BFS}. }
\begin{eqnarray}\label{U1}
U(g,f) = \nu_1 \sqrt{f}(g_{\mu\nu}-f_{\mu\nu})(g_{\alpha\beta}-f_{\alpha\beta}) ( f^{\mu\alpha}f^{\nu\beta} - f^{\mu\nu}f^{\alpha\beta})
\end{eqnarray}
where $\nu_1$ has dimensions of mass.

Flat space $g_{\mu\nu}=f_{\mu\nu} =\eta_{\mu\nu}$ is a solution to the equations of motion.  The linear theory around this solution
\begin{equation}\label{}
g_{\mu\nu} = \eta_{\mu\nu} + h_{\mu\nu}, \ \ \ \ f_{\mu\nu} = \eta_{\mu\nu} + \rho_{\mu\nu}
\end{equation}
describes \cite{ISS} a massless graviton\footnote{Note that a massless particle is expected for all choices of $U$ because the action is diff invariant.} $h_{\mu\nu}+\sigma \rho_{\mu\nu}$ plus a massive one $h_{\mu\nu} - \rho_{\mu\nu}$ with a Pauli-Fierz mass term. The mass is equal to
\begin{equation}\label{}
m^2 =  {4\nu_1(1+\sigma) \over \sigma}
\end{equation}
and the effective Newton's constant is $G(1+\sigma)$. Stability/unitarity of the linear theory then requires the two conditions
\begin{equation}\label{cond1}
{\nu_1 \over \sigma} > 0, \ \ \ \ \ \sigma > -1.
\end{equation}

The potential (\ref{U1}) breaks the original diff$\times$diff symmetry down to the diagonal subgroup. The linear theory is nevertheless well-behaved because the mass term has the Pauli-Fierz form. [See \cite{Boulanger:2000rq} for a general non-go theorem forbidding cross interactions between $N$ gravitons preserving diff$^N$.]

The applications of bigravity to massive gravity has been extensively discussed in the literature and we shall omit here. Our main goal is to discuss the properties of black holes. Black hole thermodynamics in massive gravity (with a different action) has been studied in \cite{Capela:2011mh}.

The main goal of this paper is to study in detail the properties of black holes solutions to the action (\ref{I}). We shall argue in Sec. \ref{ISsol} that the solutions found in \cite{IshamStorey}, associated to the potential (\ref{U1}), do not have regular horizons and thus they cannot represent black holes. There exists, however, a much larger class of potentials accepting exact solutions and having a unitary/stable linear theory \cite{Damour2003,Berezhiani:2008nr}. In Sec. \ref{newU} we consider a more general potential (with two parameters) and show that regular solutions do exist.  We then compute the total energy, entropy, temperature and discuss the thermodynamical properties. We shall see that phase transitions exist when $\sigma < 0$, which is allowed by unitarity/stability.

\section{The Isham-Storey exact solution}
\label{ISsol}

Soon after the action (\ref{I}) was proposed, solutions with spherical symmetry of the form
\begin{eqnarray}
  g_{\mu\nu}dx^\mu dx^\nu &=& -h(r)dt^2 + {dr^2 \over h(r)} + r^2 d\Omega^2 \\
  f_{\mu\nu}dx^\mu dx^\nu &=&   -X(r)dt^2 + Y(r) dr^2 + 2H(r)dt dr +  k_0^2 r^2 d\Omega^2 \label{ans}
\end{eqnarray}
were discovered by Isham and Storey\footnote{Note that $k0\neq 1$ does not introduce conical singularities because these solutions are asymptotically (A)dS, and have horizons. We thank R. Mann for a useful conversation on this point.} \cite{IshamStorey}. For our purposes here is enough to mention that $h(r)$ and $X(r)$ have the Schwarzschild-AdS form
\begin{eqnarray}\label{}
h(r)  &=& 1 - {2M \over r} + \Lambda r^2 \nonumber\\
X(r) &=& {3\Delta \over 2} \left( 1 - {2m \over r} + \lambda r^2 \right).
\end{eqnarray}
$M,m$ and $\Delta$ are independent integration constants and $k_0^2 = {2 \over 3}$. The cosmological constants $\Lambda$ and $\lambda$ are combinations of $\nu_1$ and $\Delta$. See \cite{IshamStorey} for full details.

A remarkable property of the solution, already noticed in \cite{IshamStorey}, is that $f_{\mu\nu}$ can also be brought to a Schwarzschild form via a coordinate change $dt = dt' + \Omega(r)dr$, with a suitable $\Omega(r)$. In the new coordinate system the metric $f_{\mu\nu}$ takes the form,
\begin{equation}\label{}
f_{\mu\nu}dx^\mu dx^\nu = -X(r)dt'^2 + {A\,  dr^2 \over X(r)} + {2 \over 3}r^2 d\Omega^2
\end{equation}
where $A$ is a constant. The zero's of $X(r)$ then represent horizons for the metric $f_{\mu\nu}$. The full solution is then simply two Schwarzschild metrics written in different coordinate system.

The Isham-Storey solution fulfills another remarkable property. Consider an arbitrary linear combination of both metrics
\begin{equation}\label{}
q_{\mu\nu} = a\, g_{\mu\nu} + b \, f_{\mu\nu}.
\end{equation}
Since $f_{\mu\nu}$ is off-diagonal,  the metric $q_{\mu\nu}$ is also off-diagonal. One can introduce a new time coordinate $t'$ defined by $dt=dt' + \Omega'(r) dr$, and again, for a suitable choice of $\Omega'(r)$ the metric $q_{\mu\nu}$ becomes diagonal. The remarkable property is that for {\it any} choice of $a,b$ this metric again has the Schwarzschild form
\begin{equation}\label{ab}
{1 \over A} \, q_{\mu\nu}dx^\mu dx^\nu = - \left( 1 - {2M' \over r} + \Lambda' r^2 \right) dt'^2 + {dr^2 \over 1 - {2M' \over r} + \Lambda'r^2   } + B\, r^2 d\Omega^2
\end{equation}
where $A,B$ are constants that depend on $A,B$, and $M',\Lambda'$ depend on all parameters $M,m,\Delta,a,b$. This metric represents a black hole with a new mass parameters that depends on the particular linear combination. In particular, the location of the horizon depends on $a,b$. Thus, for fixed values of $M,m,\Delta$, the Isham-Storey solution generates a whole family of black holes with horizons at arbitrary locations.

Summarizing, at first sight, the Isham-Storey configuration has (at least) two horizons -one for each metric- defined by the points where the functions $h(r)$ and $X(r)$ vanish.  These (candidate) horizons are located at different and independent points in spacetime because the zeroes of $h(r)$ and $X(r)$ are defined by independent integrations constants $M$ and $m$. This interpretation is however not correct.
The Isham-Storey configuration is singular because, as we now show, there is no coordinate system where both horizons can be made regular simultaneously.

As a first attempt to prove regularity, the metric $g_{\mu\nu}$ can be put in Eddington-Flinkestein coordinates \cite{Blas2006} making it regular at its own horizon. Remarkably, the metric $f_{\mu\nu}$ becomes regular at that point as well. However, this can only be achieved for either ingoing or outgoing coordinates (depending on the choice of sign for $H(r)$), but not for both simultaneously. As it is well-known \cite{Misner:1974qy}, only half of the Eddington-Flinkestein coordinates is not enough to declare regularity.

Let us concentrate on the $g_{\mu\nu}$ horizon defined by the condition \begin{equation}\label{}
h(r_g)=1 - {2M \over r_g} + \Lambda r_g^2 = 0.
\end{equation}
If $\Lambda > 0$ this equation has more than one positive solution. The argument that follows applies to all regular horizons. We assume, however, that the horizon is non-extremal, that is $h'(r_g)\neq 0$. The argument can be generalized to extreme horizons as well.  Defining a proper radial coordinate $d\rho^2 = {dr^2 \over h(r)}$ in a neighborhood of the horizon, the metric $g_{\mu\nu}$ can be brought to the form
\begin{equation}\label{a}
ds^2 = - a^2 \rho^2 dt^2 + d\rho^2 + \mbox{angular part}
\end{equation}
where $a$ is a constant. The 1-form $dt$ is singular at the horizon (now located at $\rho\rightarrow 0$) but $\rho\, dt$ is regular. The metric (\ref{a}) represents a regular spacetime written in singular hyperbolic polar coordinates.

Let us now look at the metric $f_{\mu\nu}$ in the neighborhood of $r_g$. The functions $X,Y,H$ take finite values at that point and thus, naively, the metric $f_{\mu\nu}$ looks regular,
\begin{equation}\label{fhor}
df^2 = -x\, dt^2 + y\, dr^2 + 2h \, dt dr + \mbox{angular part}.
\end{equation}
Here $x=X(r_g),\, y=Y(r_g)$ and $h=H(r_g)$ are non-zero for generic values of $M,m$ and $\Delta$. Then, since the 1-form $dt$ is singular at $r=r_g$ (or $\rho=0$), the metric (\ref{fhor}) is actually singular there. A more geometrical way to see this is to note that the set of points $(r=r_g,-\infty<t<\infty)$ with fixed angular variables correspond to a timelike one-dimensional curve in the metric $f_{\mu\nu}$, but  a 0-dimensional point in the metric $g_{\mu\nu}$. This means that metric $g_{\mu\nu}$ requires that the above set of points should be identified, while metric $f_{\mu\nu}$ require them to be all independent. The only way to make this metric regular is to assume that $X(r)$ and $H(r)$ also vanish at $r=r_g$ (with appropriate weights). A more explicit way to reach to the same conclusion is to put the metric (\ref{a}) in regular  Cartesian coordinates. This can always be done in a neighborhood of its horizon. Then one looks at the metric $f_{\mu\nu}$ in this (regular) coordinate system. It follows that $f_{\mu\nu}$ can be regular at the horizon of $g_{\mu\nu}$ if and only if its own horizon is also located there.

In summary, if $r_g$ and $r_f$ denote, respectively, the solutions to $h(r_g)=0$ and $X(r_f)=0$ we have concluded that $g_{\mu\nu}$ is singular at $r_f$, and $f_{\mu\nu}$ is singular at $r_{g}$. The only truly regular solution is one with $r_g=r_f$. This solution does exist, but corresponds to a proportional case $f_{\mu\nu}=B \, g_{\mu\nu}$ ($B$ is a constant), where $g_{\mu\nu}$ is the Schwarzschild metric.  This particular solution is the only regular black hole within the family of Isham-Storey solutions.

It is instructive to recall the Reissner-Nordstrom black hole system having a non-zero Coulomb field $A_\mu dx^\mu = A_0 dt$. The same arguments exhibited above imply that the potential $A_0(r)$ must be zero at the horizon if one demands the 1-form $A_\mu dx^\mu$ to be regular there. For the Reissner-Nordstrom black hole, this can always be achieved via a suitable gauge transformation.

The Isham-Storey exact solution associated to the potential (\ref{U1}) cannot be interpreted as a black hole, but only as the exterior solution to some mass distribution in bigravity.

\section{A bigravity black hole}
\label{newU}

We were motivated by the question of whether or not the singularity at the horizon was  a generic property of bigravity, or a property of the particular Isham-Salam-Strathdee potential (\ref{U1}).  The potential (\ref{U1}) is one member within the infinite dimensional family of potentials \cite{DamourKogan} given rise to a unitary and stable linear Pauli-Fierz theory. It is then natural to ask whether or not other potentials may give rise to regular black holes. To answer this question (in a simple way without trying to prove a general theorem) we shall consider the 2-parameter family of potentials,
\begin{eqnarray}\label{U2}
U(g,f) &=& \sqrt{f}( g_{\mu\nu}-f_{\mu\nu}) (g_{\alpha\beta}-f_{\alpha\beta} )\times \nonumber\\
&& \ \ \ \ \ \   \Big[\nu_1 ( f^{\mu\alpha} f^{\nu\beta} - f^{\mu\nu}f^{\alpha\beta}) + \nu_2( g^{\mu\alpha}g^{\nu\beta} - g^{\mu\nu}g^{\alpha\beta})\Big].
\end{eqnarray}
Note that for all $\nu_2$, this potential belongs to the Pauli-Fierz class as defined in \cite{DamourKogan}, and that for $\nu_2=0$ we recover the potential (\ref{U1}). See \cite{Berezhiani:2008nr} for a much larger class of potentials admitting exact solutions.  We shall see that, for generic values of $\nu_2$, regular, non-trivial (i.e., non-proportional) black hole solutions do exist. Since the condition $\nu_2\neq 0$ represents an open set one can conjecture that, generically, the Pauli-Fierz family of potentials does accept regular black holes solutions.

Before displaying the exact solution to the potential (\ref{U2}) we mention that (i) the potential (\ref{U2}) also accepts the background $g_{\mu\nu}=f_{\mu\nu}=\eta_{\mu\nu}$. The linear spectrum contains a massless field plus a Pauli-Fierz massive theory. The mass is now,
\begin{equation}\label{}
m^2 = {4 (\nu_1+\nu_2)(1+\sigma) \over \sigma}.
\end{equation}
As in the previous case, $1+\sigma$ must be positive to avoid ghosts. Then, linear  unitarity/stability holds for
\begin{equation}\label{un2}
{\nu_1+\nu_2\over \sigma}>0, \ \ \ \ \ \sigma> -1.
\end{equation}

Our goal is to prove that a regular solutions exist, not to classify all solutions, which is a hard task. We consider the family of potentials where $\nu_1$ is positive while $\nu_2$ is negative. We write them in the form,
\begin{equation}\label{}
\nu_1 = p^2, \ \ \ \ \nu_2=-p'\,^2.
\end{equation}

The metric components are more complicated albeit exact \cite{Berezhiani:2008nr}. The metric ansatz is again given by (\ref{ans}), and the functions $f,X,Y,H$ are now  given by the following expressions,
\begin{eqnarray}
  h(r) &=& 1  - {2M \over r} - \Lambda\,r^2 - {\sigma k_0^3 \kappa Q \over r^\alpha}  \nonumber\\
  X(r) &=& {1 \over k_0^4 \kappa^2} \left( 1 - {2m \over r} - \lambda\, r^2   + \  {Q\kappa^2 k_0^4 \over r^\alpha } \right).
   \label{sol}
\end{eqnarray}
The exponent $\alpha$ and the cosmological constants are given by
\begin{eqnarray}\label{alpha}
\alpha &=& {2(3k_0^4-4k_0^2+3)  \over (3-2k_0^2)(3k_0^2-2)  } \\
\Lambda &=& {p^2(k_0^2-1)(2k_0^8 \kappa^2-3k_0^6\kappa^2+3 k_0^2-2)  \over (2k_0^2 - 3)k_0^3\kappa }\, \nonumber\\
\lambda &=&  {p^2(k_0^2-1) (5k_0^8\kappa^2-11k_0^6\kappa^2+4k_0^4+4k_0^4\kappa^2+k_0^2-3)   \over 2\sigma  k_0^2(3-2k_0^2)  } \nonumber
\end{eqnarray}
Here $\kappa$ is a new constant defined by
\begin{equation}\label{}
\kappa = {p' \over p} \sqrt{ {3-2k_0^2 \over 3k_0^2-2 }}
\end{equation}
Unlike the Isham-Storey solution, now the parameter $k_0$ entering in the ansatz (\ref{ans}) is an arbitrary constant.

Note that this solution describes (anti)-de Sitter solutions even though there is no cosmological constant in the action. The main difference with respect to the Isham-Storey solution is the new term ${Q \over r^{\alpha}}$. As shown in \cite{Berezhiani:2008nr}, these terms are generic and appear for a large class of potentials.

Given the functions $h(r)$ and $X(r)$ above, the function $Y(r)$ is most easily expressed as the solution to the following simple relation
\begin{equation}\label{}
h\,Y+ {X \over h} = {1 + k_0^6\kappa^2 \over k_0^4 \kappa^2} + {3(4-7k_0^2+4k_0^4)  \sigma k_0^2 \over p^2(2k_0^2-3)(3k_0^2-2)^2 } \, {Q \over r^{\alpha+2}}
\end{equation}
Finally, knowing $X(r)$ and $Y(r)$, the function $H(r)$ is given by
\begin{equation}\label{}
H(r) = \pm \sqrt{ {1 \over k_0^2\kappa^2} - X(r)Y(r)  }
\end{equation}
These relations fix completely all unknown functions. We then have an exact solution to the equations of motion following from the action (\ref{I}) with the potential (\ref{U2}).

\begin{figure}[htb]
\centerline{\includegraphics[scale=0.6]{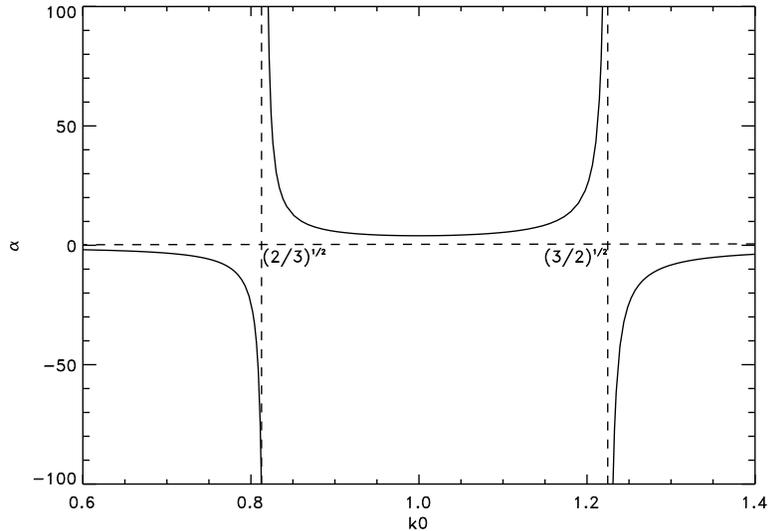}}
\caption{Plot of $\alpha(k_0)$. In the range $\sqrt{2/3} < k_0 < \sqrt{3/2}$ the exponent $\alpha$ is always positive and greater than $4$.\label{alpha.fig}}
\end{figure}

Contrary to the Isham-Storey solution, the parameter $k_{0}$ entering in the ansatz (\ref{ans}) is now an arbitrary integration constant.  The range of this constant is severely restricted if we demand the charge $Q$ to be localized. Since $Q$ enters in the metric in the form ${Q \over r^{\alpha}}$ where $\alpha$ is given in (\ref{alpha}). Demanding $\alpha$ to be positive requires
\begin{equation}\label{k0}
\sqrt{{2 \over 3}} < k_0 < \sqrt{{3 \over 2}}.
\end{equation}
A plot of $\alpha(k_0)$ is displayed in Fig. (\ref{alpha.fig}).

\begin{figure}[htb]
\centerline{\includegraphics[scale=0.6]{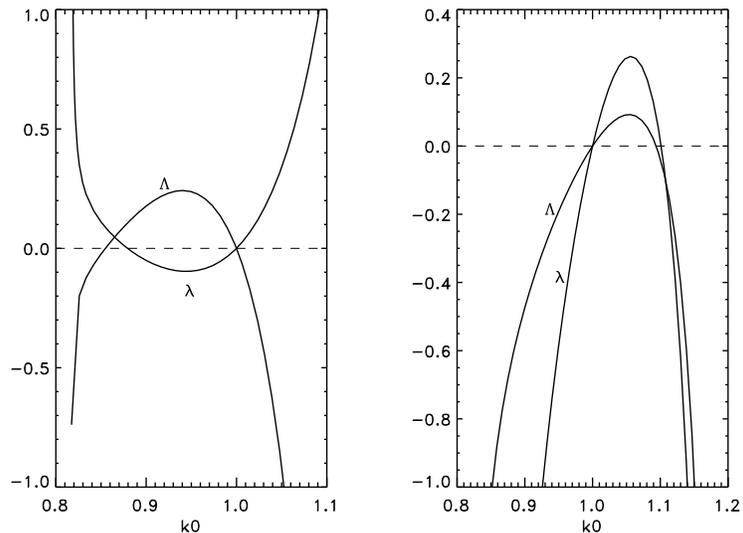}}
\caption{Plot of $\Lambda(k_0)$ and $\lambda(k_0)$. Left panel $\sigma>0$; right panel $\sigma<0$.
\label{cosm}}
\end{figure}

The cosmological constants appearing in our exact solution are not fundamental constants but depend on the state (see (\ref{alpha})). We plot in Fig. \ref{cosm} the values of $\Lambda$ and $\lambda$ for the allowed range of $k_0$ given in (\ref{k0}). We see that de Sitter and anti-de Sitter phase may coexist.  We refer the reader to \cite{Gomberoff:2003ea} for a variational formulation on de Sitter spaces. Note that for $\sigma<0$, which will turn out to be the most interesting case, both cosmological constants are negative for most values of $k_{0}$, with a small de Sitter window.

The main property of this solution is the appearance of 4 integration constants  $M,m,Q$ and $k_0$. One would like to give an interpretation to these constants in terms of asymptotic conserved charges. This is however not possible. As has been extensively discussed in the literature (see, for example, \cite{BlasDeffayetGarriga}), bigravity solutions break Lorentz (or de Sitter) invariance. In fact, in our case, the only asymptotic Killing vector is constant time translations.  In this sense there is only one conserved charge at infinity.

Interestingly, black holes do indeed have only one charge. As we see in next section regularity of the horizon imposes three restrictions on the values of $M,m,Q,k_0$. As a result, there is only one remaining free parameter that we can identify as the total energy (conjugate to time translations).

\section{Regularity conditions}

The horizon for the metric $g_{\mu\nu}$ is given by the inner solution of the equation $h(r)=0$. (The outer solution, if any, will be a cosmological horizon). As before, the metric $f_{\mu\nu}$ has its own horizon defined by the equation $X(r)=0$.  As discussed before, the only way to have a regular horizon is to ensure that both $X(r)$ and $h(r)$ vanish at the same point. Moreover, the function $H(r)$ -which appears in the metric multiplied by the singular 1-form $dt$- has to vanish at the horizon as well.

There is one more restriction to achieve regularity. The ``temperatures" of both metrics must be the same. This is a condition in either the Minkowskian or Euclidean sectors. Kruskal coordinates will exists simultaneously for both metrics provided the derivatives $f'(r)$ and $X'(r)$ evaluated at $r=r_g$ are correlated. This is most easily seen in the Euclidean sector as a condition eliminating a conical singularity. But it is equally valid in the Minkowskian sector as a condition for the Kruskal extension to exist.

Summarizing, let $r_{g}$ the zero of $h(r)$,
 \begin{equation}\label{h}
 h(r_g)=0.
 \end{equation}
Recall that $h(r)$ is given in (\ref{sol}). This equation merely defines $r_g$ without really imposing any restriction.

Now, we demand the horizon of the $f_{\mu\nu}$ metric to be located at the same spacetime point, that is
\begin{equation}\label{X}
X(r_g)=0.
\end{equation}
Next, as discussed above, we also demand the function $H(r)$ to vanish at this point
\begin{equation}\label{H}
H(r_g)=0.
\end{equation}
Finally, the temperatures of both horizons must be the same so that a Kruskal extension exists for both simultaneously. This condition reads
\begin{equation}\label{T}
f'(r)|_{r_g} = x_0 X'(r)|_{r_g}
\end{equation}
where $x_0$ is a complicated constant that we prefer to omit.

Ideally, one would like to use (\ref{h},\ref{X},\ref{H},\ref{T}) to express, for example, $M,m,Q,k_0$ in terms of $r_g$. This is algebraically impossible. What we can do, it to use these equations to write $M,m,Q$ as functions of $k_0$. Thus, a regular black hole is built by choosing a value of $k_{0}$, within the allowed range (\ref{k0}). Once $k_{0}$ is given, all other charges
\begin{equation}\label{charges}
M(k_0), m(k_0),Q(k_0)
\end{equation}
take definite values. The explicit formulas are analytical but not very illuminating so we skip them. It is more illustrative to present plots of $M(k_0),m(k_0),Q(k_0)$ for the allowed range of $k_0$.

\begin{figure}[htb]
\centerline{\includegraphics[scale=0.6]{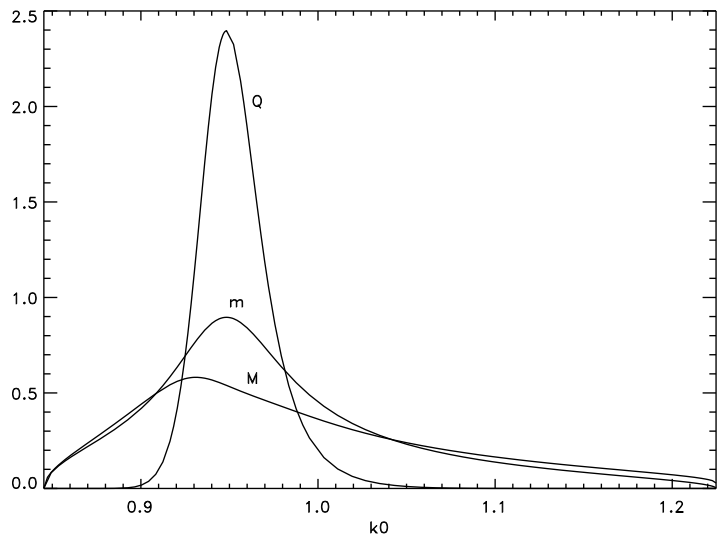}}
\caption{$\sigma = 0.5$ $p=1$, $p'=0.5$  }\label{Sigmapos.fig}
\end{figure}

\begin{figure}[htb]
\centerline{\includegraphics[scale=0.6]{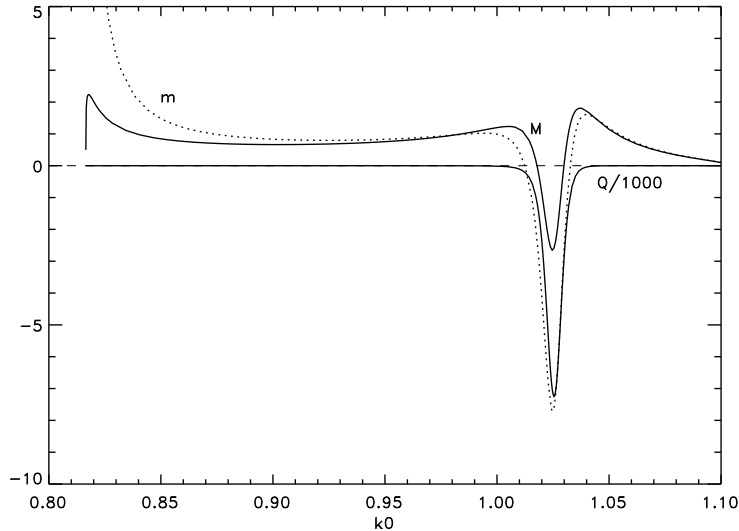}}
\caption{$\sigma = -0.7$ $p=1$, $p'=1.28$  }\label{Sigmaneg.fig}
\end{figure}

Fig. (\ref{Sigmapos.fig}) shows $M,m,Q$ for $\sigma$ positive and Fig. (\ref{Sigmaneg.fig}) shows $M,m,Q$ for $\sigma$ negative. These pictures are generic and do not change drastically for small variations of the parameters within the allowed ranges.

~

One may be concerned that imposing all these conditions require severe fine tuning and the solution is highly non-generic.  Note, however, that this fine tuning must not be imposed. The parameters $Q,m, M$ of the solutions are not conserved quantities. Only one particular combination of them, the total energy, which will be discussed in the next section, is conserved during gravitational collapse. The particular way the collapse takes place will not be considered here, but it is a sensible guess, that in order to satisfy  the cosmic censorship conjecture, the parameters will evolve to an equilibrium value that fulfill the constraint. From the canonical point of view we expect that this may also be derived from an extremum principle, in the same way that the constraint relating energy and temperature may be derived in the Euclidean treatment of black hole thermodynamics. We leave this to future work.

\section{Total energy and black hole temperature}

We now turn to the problem of computing the mass, and thermodynamical functions for this black hole.

Ideally, we should set up boundary conditions appropriated to the black hole problem, preserving some asymptotic group. However, bigravity solutions are neither asymptotically flat nor (anti-)de Sitter. For the black holes displayed above, each metric is asymptotically (anti-) de Sitter but with different speeds of light. There is no obvious asymptotic structure one would like to preserve. In order to move forward we do the calculation in the quickest way by allowing in the variational principle at least the family of metrics described by eh black hole, varying all parameters.

Since the potential in the action has no derivatives of the fields, it does not contribute to the boundary terms. The total energy is clearly given by two ADM functionals, one for each metric. By usual methods \cite{ReggeTeitelboim}, we find for the variation of the Hamiltonian,
\begin{eqnarray}\label{}
\delta E &=& \delta E(g) + \delta E(f) \nonumber \\
        &=& \delta M - {5p'\, m\, k_0^2 (6k_0^4-11k_0^2+6) \delta k_0 \over p(3-2k_0^2)^{1/2} (3k_0^2-2)^{3/2} } \label{dE}
\end{eqnarray}
plus terms that vanish on-shell. As explained above, the second line follows by plugging the black hole solution in the variation of the boundary term, varying all parameters.  $\delta M$ is a expected contribution from $E(g)$ while the second term comes from $E(f)$.

It is very interesting to observe  that both $\delta E(g)$ and $\delta E(f)$ have divergent contributions (as $r\rightarrow\infty$).  These are expected because the metrics have terms of the form $\Lambda r^2$ where $\Lambda$ is not a fundamental constant. Instead, $\Lambda$ depends on integration constants (see (\ref{alpha})) which are varied in the action principle. However, these divergent pieces nicely cancel each other and the sum $\delta E(g)+ \delta E(f)$ remains finite. A phenomenon like this one has been observed in other systems \cite{Henneaux:2002wm}. A general argument has been given in \cite{Barnich:2001jy}.

Note that the second term depends on $m$ and $k_0$. One may conclude that this solution has a new asymptotic symplectic pair $k_0,m$ (a charge with a corresponding chemical potential). This interpretation is however not correct. There is only one asymptotic time translation symmetry and only the total energy is a conserved charge.

Integrability of the total energy then implies that $m$ cannot be varied arbitrarily, but must be a function of $k_0$. We have already seen that $m$ is related to $k_{0}$ for regular black holes. We  use this relation and write the total energy as
\begin{equation}\label{E}
E = M(k_0) - {5p' \over p} \,\int^{k_{0}}_1 dk_0 { m(k_0)\, k_0^2 (6k_0^4-11k_0^2+6) \over (3-2k_0^2)^{1/2} (3k_0^2-2)^{3/2} }
\end{equation}
where $m(k_0)$ is plotted in Figs. \ref{Sigmapos.fig} and \ref{Sigmaneg.fig} as function of $k_0$. We have arbitrarily chosen $k_0=1$ as a reference state. This is an allowed value and corresponds to the point where the exponent $\alpha(k_0)$ takes its minimum value $\alpha(1)=4$.

~

The temperature of the black hole can be computed either from the $g_{\mu\nu}$ or $f_{\mu\nu}$ black hole. Applying the usual formula for the $g_{\mu\nu}$ black hole,
\begin{equation}\label{T}
T = {1 \over 4\pi} \left .{dh(r) \over dr}\right|_{r_g}.
\end{equation}

\begin{figure}[htb]
\centerline{\includegraphics[scale=0.6]{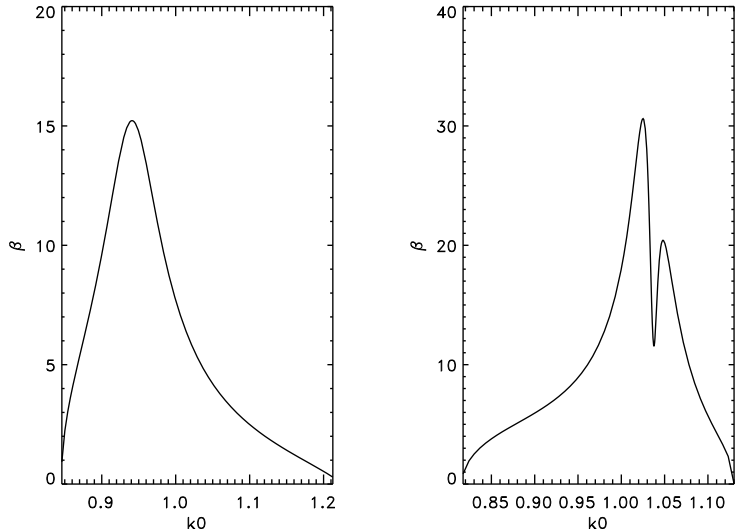}}
\caption{Left panel: $\sigma=0.5,p=1, p2=0.5$. In this case, there are two states (two values of $k_0$) for each temperature. Right panel: $\sigma=-0.7,p=1, p2=1.405$. Here there are up to four states for each temperature.  This indicates possible phase transitions. The temperature $T$ diverges near the extreme allowed values $k_0^2=2/3$ and $k_0^2=3/2$. The horizon also vanishes at those points and the behavior resembles that of Schwarzschild. }\label{beta}
\end{figure}

In Fig. \ref{beta} we plot the temperature for two choices of parameters $\sigma,p$ and $p'$. Focus first on the left panel with $\sigma=0.5,p=1, p2=0.5$.  We see that for a given value of $\beta$ (not too large) there exists two values of $k_0$, and hence two values of $M,m,Q$. In the canonical ensemble, where $\beta$ is fixed, there exists two black holes consistent with that temperature. To discern which one is realized one needs to compute the free energy. For negative values of $\sigma$ the situation is even more interesting. On the right panel we plot $\beta(k_0)$ for $\sigma=-0.7,p=1, p2=1.405$. For a given value of $\beta$ there are up to 4 values of $k_0$ and thus 4 states. Again, to discern which one is the most stable one we need to compute the free energy.  The free energy will also tell us whether or not phase transitions among these states can occur or not. We do this calculation in the next section.

\section{Entropy and free energy}

As shown in Fig. (\ref{beta}), for a given temperature there can be up to 4 values of $k_0$. Hence, for a given temperature there may be up to four different black hole metrics. To decide which is thermodynamically favored we need to compute the free energy $F = E-TS$.

The logic here mimics step by step the usual GR calculation, although the details are more involved. The full Hamiltonian action must be supplemented by two boundary terms at the horizon and infinity,
\begin{equation}\label{If}
I = \int \mbox{(bulk)} \ + \   \beta\, E - S(r_g)
\end{equation}
where $E$ is given in (\ref{E}) and ($G=1$)
\begin{eqnarray}\label{}
S(r_g) &=& \pi r_g^2  + \sigma \pi k_0^2 r_g^2 \nonumber\\
 &=&  {A(g) \over 4} + \sigma {A(f) \over 4}.
\end{eqnarray}
This formula for the horizon boundary term follows from general grounds (it can also be derived straightforwardly). Indeed, the full action is a sum of Einstein-Hilbert actions. Therefore, the total entropy is expected to be a sum of the individual Bekenstein-Hawking entropies. Note that the second action is multiplied by $\sigma$, hence the second entropy is multiplied by $\sigma$. Up to this point, the horizon $r_g$ has no logical relation with the asymptotic charges $M,m$, etc. The action principle (\ref{If}) is appropriated to black hole fields having a horizon at some arbitrary point $r_g$.

The next step is to evaluate the action on a saddle-point approximation. This has two implications.  First, we solve the constraints plugging the regular black hole solution. The bulk pieces vanish (because there are a combination of constraints) and the total action becomes
\begin{equation}\label{}
I = \beta E - S(E)
\end{equation}
where now the entropy does depend on the asymptotic charge. This is not the end yet. Since $E$ is varied ($\beta$ is fixed),  there is one remaining equation $\delta I/\delta E=0$ which implies the first law:
\begin{equation}\label{entropy}
\beta = {\delta S \over \delta E}.
\end{equation}
Of course, one can check that this value for $\beta$ is consistent with the `no-cone' condition (\ref{T}).

The free energy ${1 \over \beta} I = E - TS=F(\beta)$ is a function of the temperature. Let us now evaluate  $F(\beta)$ for various cases. As before, the formulas are analytic but not very illuminating. We display the main results with plots. Note that both $F(k_0)$ and $\beta(k_0)$ (for  all allowed $k_0$) are known and hence we can plot $F(\beta)$ parametrically.

We shall vary the couplings $\sigma,p$ and $p'$ in a way consistent with stability/unitarity of the linear theory, as discussed in Sec. \ref{newU}.  For given values of $\sigma,p,p'$ we then vary $k_0$ within the range (\ref{k0}).

\begin{figure}[htb]
\centerline{\includegraphics[scale=0.6]{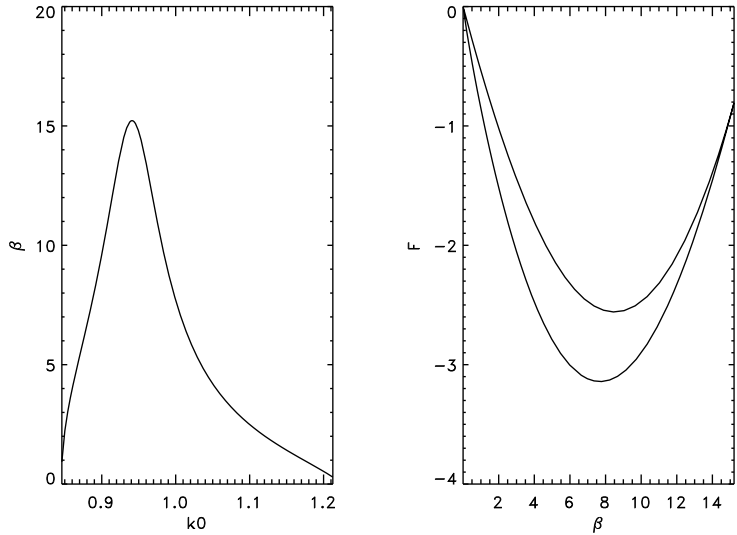}}
\caption{$\sigma=0.5$ and $p=1,p'=0.5$}
\label{s+}
\end{figure}

First, consider the case $\sigma=0.5,p=1,p'=0.5$. Recall that for $\sigma$ positive $p$ must be bigger than $p'$. The right panel in Fig. \ref{s+} shows  $F(\beta)$. We recall that for a given value of $\beta$ there are two values of $k_0$ (left panel) and thus two allowed states. As a consequence, the free energy is a multi-valued function of $\beta$ and the lowest value determines the most stable state. In this case, the lowest $F$ corresponds to the value of $\beta$ with the biggest $k_0$. This is true in the whole range of $k_0$ and hence there are no phase transitions  in this case.

\begin{figure}[htb]
\centerline{\includegraphics[scale=0.6]{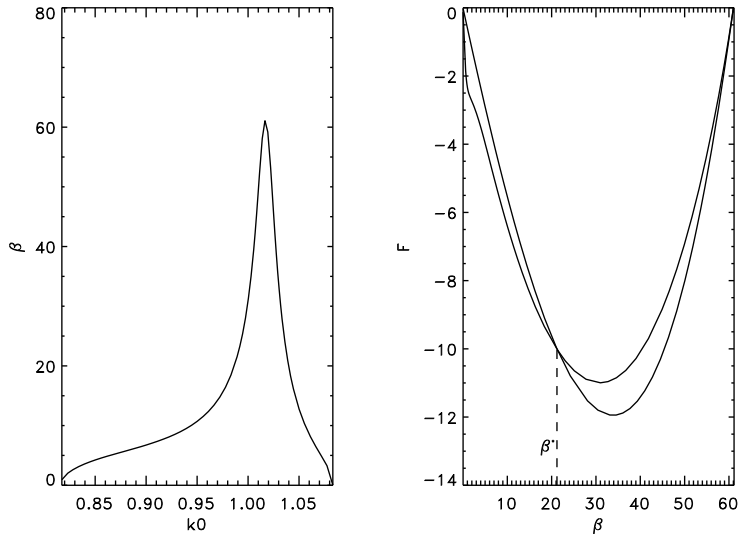}}
\caption{$\sigma=-0.7$ and $p=1,p'=1.2$}
\label{s-1}
\end{figure}

A more interesting situation occurs for  $\sigma=-0.7,p=1,p'=1.2$ (for $\sigma<0$, linear unitarity/stability requires $p'>p$).  We plot in Fig. \ref{s-1} $F(\beta)$. As before, there exists two states for a given $\beta$. However, in this case, there exists a phase transition for $\beta = \beta^*$. For $\beta>\beta^*$, the most stable state is the one with biggest $k_0$, while for $\beta<\beta^*$, the most stable state is the one with smallest $k_0$. The physical free energy is the envelope of the curve $F(\beta)$ with a discontinuous derivative at $\beta=\beta^*$. Hence, a first order phase transition takes place at that temperature.

\begin{figure}[htb]
\centerline{\includegraphics[scale=0.6]{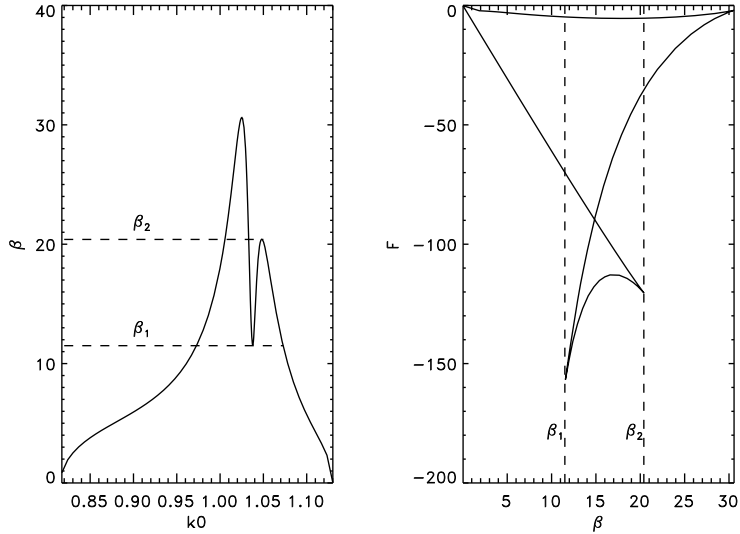}}
\caption{$\sigma=-0.7$ and $p=1,p'=1.405$}
\label{s-2}
\end{figure}

An even richer situation arises if we increase the value of $p'$, leaving $p$ fixed. For $\sigma=-0.7,p=1,p'=1.405$ we plot in Fig. \ref{s-2} the temperature and $F(\beta)$. For the range $\beta_1<\beta <\beta_2$, four states are available for a given $\beta$. The free energy $F(\beta)$ (right panel) also exhibits this degeneracy and allows to determine the most stable state: for $\beta<\beta_1$ and $\beta>\beta_2$ corresponds to the the biggest value of $k_0$, and for $\beta_1<\beta<\beta_2$ corresponds to the second biggest value of $k_0$.\\
In this case, there appears to be a ``zero-order" phase transition when crossing the critical values $\beta=\beta_1$ and $\beta=\beta_2$. This hypothetical zero-order phase transitions has been discussed before in the context of self-gravitating gases (see \cite{deVega:2001zk} and \cite{PhysRevE.65.056123} for details).

\begin{figure}[htb]
\centerline{\includegraphics[scale=0.6]{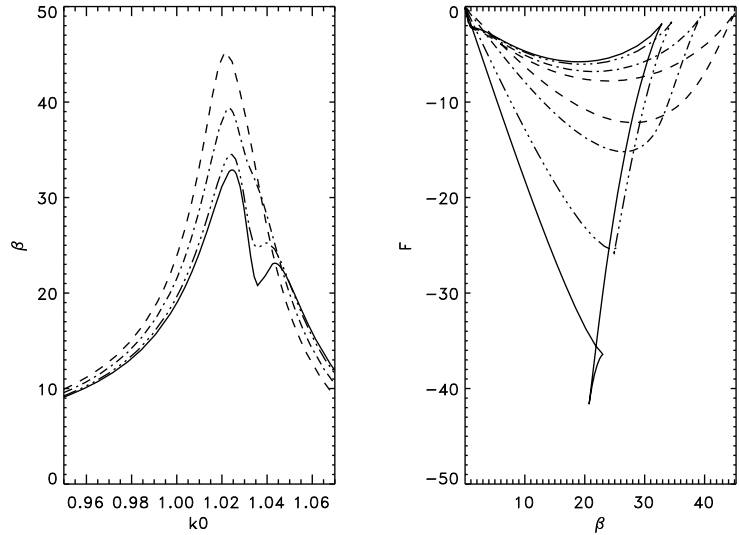}}
\caption{$\sigma=-0.7,p=1$ and $p'=\{1.282,1.323,1.364,1.38\}$}
\label{s-3}
\end{figure}

Finally, in Fig. \ref{s-3} we plot  the temperature and free energy for a sequence of values of $p'$, leaving $p=1$ fixed showing how the systems moves from two to four states. Since $p'$ is a coupling constant in the action we would not interpret this effect as a second order transition.

\section{Conclusions}

To conclude, we have argued that the Isham-Storey exact solution cannot represent a black hole but the exterior solution to a spherical star. This is not a generic property of bigravity, but of  the potential chosen by Isham and Storey. For more generic potentials we find more integration constants in the solutions, which may be fine tuned to produce a regular horizon.
We have computed the total energy, temperature and free energy of these solutions and have found several phases in the canonical formalism. Phase transitions do occur for certain critical values of the temperature, depending on the coupling $\sigma$ and the potential parameters $p$ and $p'$.

\section{Acknowledgments}

MB was partially supported by Fondecyt (Chile) Grants \#1100282 and \# 1090753. The work of AG was partially supported by Fondecyt (Chile) Grant  \# 1090753. MP was supported by CONICYT grant and would like to thank P. Arriagada for useful tips.


\end{document}